\documentstyle[aps, prl, twocolumn, floats, epsfig]{revtex}

\title{    Simulations on Infinite Size Lattices}
\author{   H.G. Evertz, W. von der Linden}
\address{  Institut f\"ur Theoretische Physik, Technische Universit\"at Graz, 8010 Graz, Austria}
\date{April 5, 2001}

\newif\ifdraft

  \ifdraft
   \newcommand{\note }[1]{{\bf [{ #1~}]}\marginpar{~~~~~~~\rule[0ex]{0.5mm}{2ex}} }
    \newcommand{\notel}[1]{\note{#1}\\}
  \else
    \newcommand{  \note }[1]{}
    \newcommand{  \notel}[1]{}
  \fi

\newcommand{\beq}[1]{\begin{equation}\label{#1}}
\newcommand{\eeq}{\end{equation}}
\newcommand{\beqa}[1]{\beq{#1}\begin{array}{lllllllll}}
\newcommand{\eeqa}{\end{array}\eeq}
\newcommand{\beqar}[1]{\begin{eqnarray}\label{#1}}
\newcommand{\eeqar}{\end{eqnarray}}
\newcommand{\eq}[1]{eq.\ (\ref{#1})}       
\newcommand{\fig}[1]{fig.\ \ref{#1}}       
\newcommand{\Fig}[1]{Fig.\ \ref{#1}}
\newcommand{\Label}[1]{\label{#1}}
\newcommand{\bit}{\begin{itemize}}
\newcommand{\eit}{\end{itemize}}

\begin{document}

\twocolumn[\hsize\textwidth\columnwidth\hsize\csname @twocolumnfalse\endcsname
\maketitle
\begin{abstract}
We introduce a Monte Carlo method,
as a modification of existing cluster algorithms,
which allows simulations directly on systems of infinite size,
and for quantum models also at $\beta=\infty$.
All two-point functions can be obtained, including dynamical information.
When the number of iterations is increased, correlation functions at larger distances
become available. 
Limits $q\to 0$ and $\omega\to 0$ can be approached directly.
As examples we calculate spectra for the d=2 Ising model 
and for Heisenberg quantum spin ladders with 2 and 4 legs.
\end{abstract}
\pacs{}
]

Standard Monte Carlo simulations are limited to systems of finite size.
Physical results for infinite systems have to be obtained by
finite size scaling, assuming that one knows the correct scaling laws,
and {assuming} that the data are already in a suitable asymptotic regime.
It is therefore very desirable to obtain results also {\em directly} 
at infinite system size.
We will show how to do so, with only a small modification 
of existing cluster algorithms, by using the cluster representation
of the models to calculate two-point functions.
In the quantum case we can then also simulate directly at $\beta=\infty$
and calculate correlation functions and dynamical greens functions.
As examples we will show calculations for the classical Ising model
and for quantum Heisenberg ladder systems.

The {\em Swendsen Wang cluster method} \cite{SW} for the classical Ising model
is based on the Edwards-Sokal-Fortuin-Kasteleyn \cite{FortuinKasteleyn,EdwardsSokal} representation 
\def\Wij{W_{ij}(s_i,s_j,b_{ij})}
\beqar{FK}
 e^{\beta J (s_i s_j \;-\;1)} &=& 
       \sum_{b_{ij}=0,1} ~p\,\delta_{s_is_j}\,\delta_{b_{ij},1} \;+\; (1-p)\,\delta_{b_{ij},0}\\
  &=:& \sum_{b_{ij}=0,1} \Wij
\eeqar
for the Boltzmann weight of a spin-pair,
with $p=1-e^{-2\beta J}$, 
which enlargens the phase space of spin variables $s_i$ by additional bond variables $b_{ij}$.
The partition function 
$Z = \sum_{\{s_i\}} \sum_{\{b_{ij}\}} W(s,b)$
with total weight $W(s,b)=\prod_{\langle ij \rangle} \Wij$
is then simulated efficiently by switching between representations:
given a spin-configuration $s \;:=\; \{s_i\}$, 
one generates a bond configuration $b \;:=\; \{b_{ij}\}$ 
with the conditional probability 
$p(b|s) \sim W(s,b)$, thus
creating a configuration of {\em clusters} of sites connected by bonds $b_{ij}=1$.
Given a bond configuration, a new spin configuration is generated with probability 
$p(s|b) \sim W(s,b)$.
Because of the factor $\delta_{s_i s_j}\delta_{b_{ij},1}$
in $W_{ij}$, this  amounts to setting all spins of each cluster 
randomly to a common new value, independent of other clusters.

Observables $\hat{O}$ can be computed either in spin-representation as $O(s)$
or in bond representation as 
$ O(b) =  
       \left( \sum_s \;O(s)\; W(s,b)\right)
    /  \left( \sum_s \;       W(s,b)\right)
$
(so called ``improved estimators'') \cite{Wolff}.
For two-point functions $O(s) \;=\; s_i s_j$
the bond representation is particularly simple:
\beq{bondcorr}
 O(b) ~=~ \delta(\mbox{sites $i$ and $j$ are in the same cluster}) ~.
\eeq
Thus two-point functions and derived quantities (including susceptibility, energy, specific heat)
can be computed from the properties of individual clusters.

A variant of the Swendsen-Wang method is 
Wolff's {\em single cluster method} \cite{Wolff}.
Given a spin configuration, only one of the bond-clusters is constructed,
namely the cluster which contains a randomly chosen intial starting site $x_0$.
All spins in this cluster are then flipped.
One advantage is that the single cluster will on average be larger than the average Swendsen-Wang cluster,
so that its flip results in a bigger move in phase space.
Using \eq{bondcorr}, the correlation function can then be measured as
\beq{I}
 C(r) =  \langle s_i \; s_{i+r} \rangle ~=~ \langle \frac{1}{V_{cl}} 
        \sum_{i \mbox{\scriptsize ~in cluster}}
      ~\delta(\mbox{$i+r $ in cluster})\rangle ~,
\eeq
where $V_{cl}$ is the number of sites in the single cluster,
and the brackets $\langle ... \rangle$ on the right refer to the Monte Carlo average.
The susceptibility is then $\chi = \beta \langle V_{cl} \rangle$,
the average energy can be measured either as $E=-J\, 2d C(1)$ or from the average number of
bonds in the cluster \cite{FortuinKasteleyn}, and the specific heat can be obtained either as a 
numerical derivative or directly from the bond representation \cite{Caselle}.


Our new {\em infinite system method} now employs the single cluster method,
except that it starts each cluster not from a randomly chose site, but always from the {\em same} site 
$x_0$ (e.g. the origin of the coordinate system).
Our method satisfies detailed balance and ergodicity (for any finite region including the origin)
similarly to Wolff's method.
For the ferromagnetic Ising model, we begin with an initial staggered spin configuration of unlimited size.
(Only a finite part if this configuration will have to be stored).
We iterate the following two steps:
(1) For the current spin configuration, a cluster containing site $x_0$ is constructed 
    with Swendsen-Wang bond-probabilities, and
(2) all spins in this cluster are flipped, resulting in a new spin configuration.
The current cluster is discarded after each iteration.

After a sufficient total number $N(r)$ of iterations, this process will ``equilibrate'' 
all spins within a radius $r$ around $x_0$,
and   two-point functions can then be measured within this area.
With increasing number of iterations, the cluster will occasionally reach larger and larger distances.
Because of \eq{I}, the probability to do so is (roughly) proportional to the
correlation function $C(r)$.
A region of the lattice at distance $r$ from $x_0$ will become ``equilibrated'' after 
the cluster has reached that region a sufficient number $n_{eq}$ of times,
thus $N(r) \sim \frac{n_{eq}}{C(r)}$.
Since the cluster can grow without bounds, we are calculating correlation functions for the 
{\em infinite} lattice,
while we only need to store the spin configuration within the area actually reached by 
a cluster during a finite run.


\Fig{Ising} shows as an example the correlation function for the two dimensional Ising model at 
$\beta = 0.42 ~<~ \beta_c = 0.44068..$,
calculated according to \eq{I}.
\begin{figure}[htb]
\center{\epsfig{file=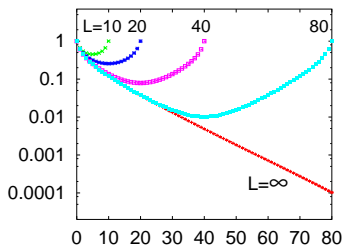,height=36mm}}  
\caption{Correlation functions of the $d=2$ Ising model at $\beta=0.42$.
}\Label{Ising} 
\end{figure}
%
For comparison we show results for finite size lattices, with periodic boundary conditions.
Usually, the infinite lattice result can only be obtained as the extrapolation of the finite lattice ones.
In contrast, with our new method we can obtain the infinite latttice result in a single simulation,
producing the lower straight line in \fig{Ising}.
We checked that the method works just as well in three dimensions.

The calculation of error bars for $C(r)$ needs special care.
For each distance $|r|$, we demand that the cluster contains pairs of sites at that distance 
$|r|^{d-1}\times n_{eq}$ times, i.e.\ it reaches
each site at that distance  $O(n_{eq})$ ($= O(10)$) times,
before we consider data at that distance thermalized. Only then do we  start to accumulate 
measurements for $C(r)$ according to \eq{I}
for all sites $i,j$ in the cluster.
Alternative similar procedures are possible.

Note that the overall computational time is by far dominated by measurements,
both in usual cluster algorithms and in our method:
In order to obtain a relative error $\epsilon$ for $C(r)$,
a large number of about $1/\epsilon^2$ clusters with sites at distance $r$ need to occur,
each of which is generated with a probability of about $1/C(r)$,
in both cases.
Thus our method is at least as efficient as usual cluster algorithms,
while providing results for the infinite lattice directly.

The average computational time {\em per cluster} is proportional to the
average cluster size, which is proportional to the susceptibility (see below \eq{I}).
The effective simulated system size grows with the number of iterations.

When will the new method work ?
The computational effort and thus the susceptibility needs to be finite.
In addition, there must not be a finite probability for a percolating cluster, 
which implies $\beta < \beta_c$.
The present implementation of the new method will thus work anywhere in the 
unbroken phase, right up to the critical point.
Systems with long range order may become accessible with a suitable modification of the method.
We note that indeed points away from the transition are often the region of
interest, when comparing results of simulations to experimental measurements.

%
%
For {\em nonrelativistic quantum systems}, the loop algorithm \cite{LoopAlg} provides a cluster method.
It is based on an enlargened representation in terms of the original spin operators and additional
loop operators, similar in spirit to the Edwards-Sokal-Fortuin-Kasteleyn representation, \eq{FK}.
We will in the following specialize to the spin $\frac{1}{2}$ quantum Heisenberg model
\beq{HB}
 {\cal H} ~=~ J \; \sum_{\langle ij \rangle} ~\frac{1}{2} \left(S_i^+S_j^- + S_i^- S_j^+ \right)
                  ~+~ \lambda S_i^z S_j^z
                  ~.
\eeq 
The two-point function for the single cluster version of the loop algorithm then reads
\beq{twopt}
  \begin{array}{llll}
  \left\langle  \frac{1}{2} \left(S_i^+ S_j^- + S_i^- S_j^+ \right) \right\rangle ~=~ 
           \left\langle      \delta(\mbox{i and j on the loop}) \right\rangle\\[1ex]
  \left\langle                    S_i^z S_j^z                       \right\rangle ~=~ 
       \left\langle S_i^z\,S_j^z ~    \delta(\mbox{i and j on the loop})\right\rangle\\
  \end{array}
   ~.
\eeq
%
Therefore our approach to simulate infinite size systems can be used in the same way as for the Ising model.
``Infinite size'' here can be applied to the spatial directions, as well as, independently, to the
direction of imaginary time, yielding simulations directly at $\beta=\infty$,
while retaining all dynamical information.
\begin{figure}[bht]
\center{
 \epsfig{file=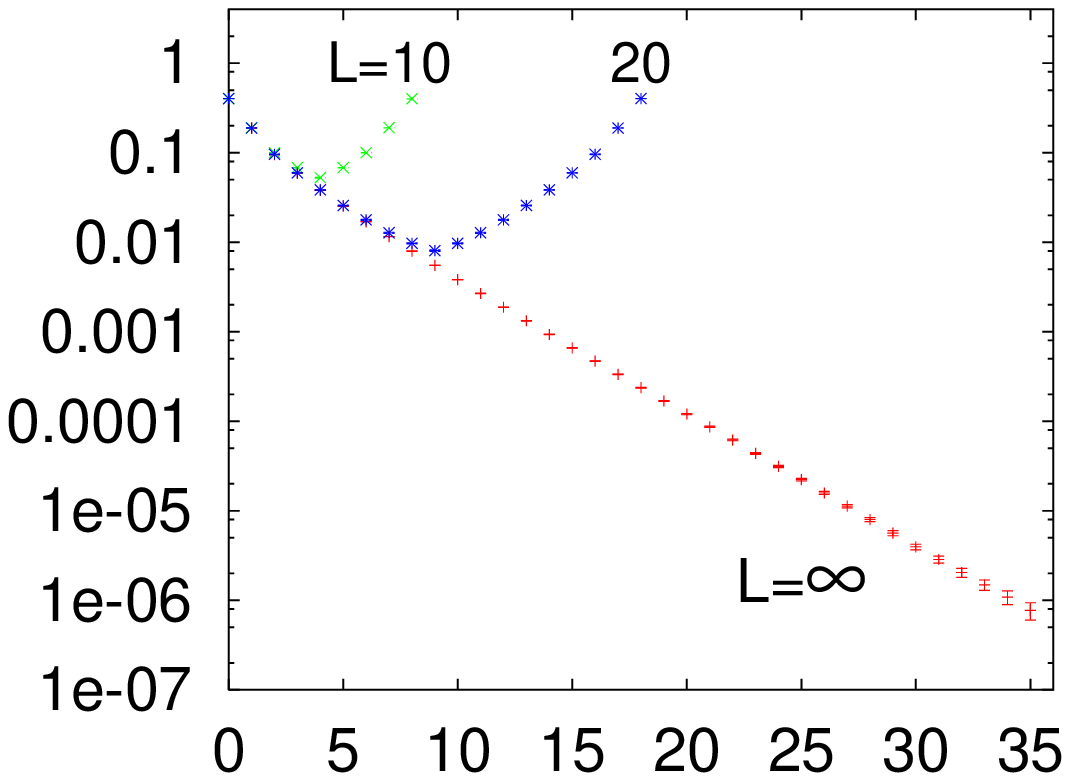,height=36mm}  
 \epsfig{file=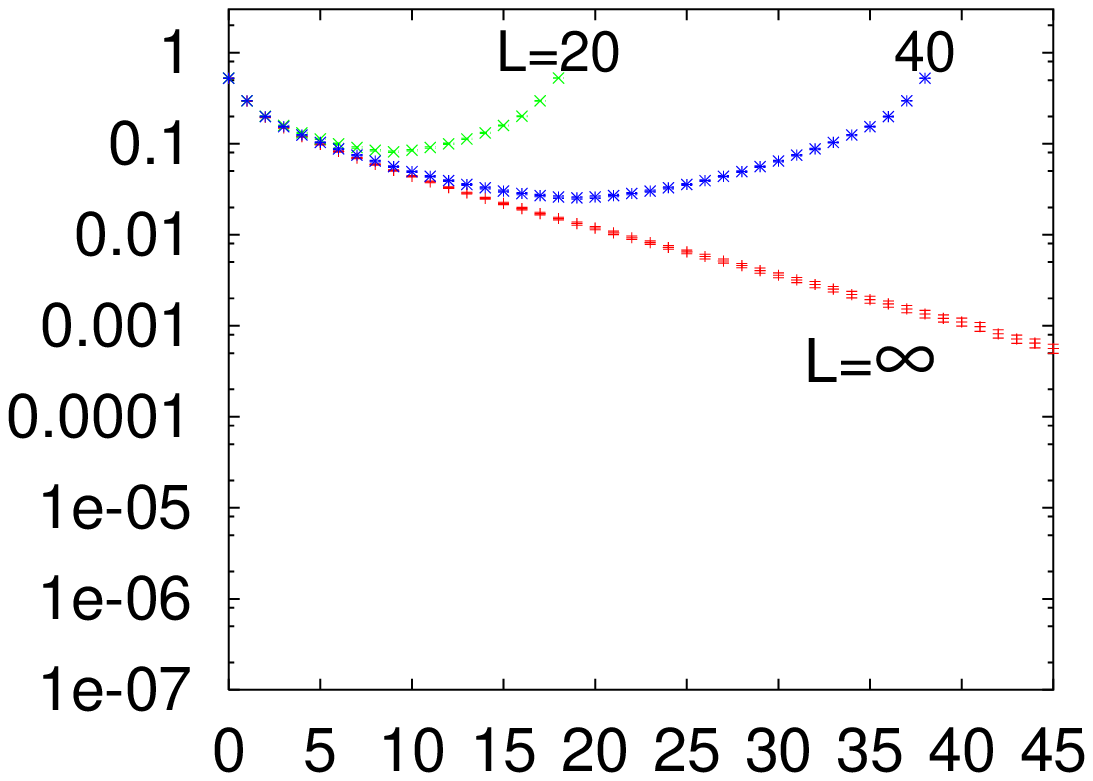,height=36mm}  
       }
     \caption{Equal time staggered spatial correlation function 
              at $q_\perp =\pi$ and $\beta=\infty$,
              for N=2 (top) and N=4 (bottom)
}\Label{spatial} 
\end{figure}
Note that our method is not a projector method. 
We can simulate directly {\em at} $L=\infty$ and $\beta=\infty$.
This also enables us to obtain the limits $q\to 0$ and $\omega\to 0$ directly from the simulations.
Contributions from $q\equiv 0$ or $\omega\equiv 0$, which are in general different from those limits, 
and are present in normal simulations with 
periodic boundary conditions, are completely avoided in our approach,
as well as the double limit $L\to\infty$ and $q\to 0$ that is usually required.
When the number of iterations  increases, clusters will reach larger distances in spatial and/or imaginary time direction,
thereby improving the  resolution in $q$ and $\omega$.

Again, our present method will be applicable provided  that 
there is no percolation and that the unsubtracted correlation function drops off
sufficiently quickly (i.e. that the corresponding susceptibility is finite).


%
As an example we studied spin ladder systems \cite{ladders} 
with $N=2$ and $4$ legs for the isotropic antiferromagnet ($\lambda=1$).
We used the discrete time version of the loop algorithm ($\Delta\tau=1/16$).
The method can be applied just as well in continuous time \cite{conttime}.
\Fig{spatial} shows 
results for 
the equal time staggered spatial correlation functions along the chains,
which behave similarly to the Ising case. 
A fit to the infinite lattice result gives $\xi=2.93(2)$ for $N=2$ and $\xi=8.2(1)$ for $N=4$. 
\Fig{temporal} shows greens functions for $L=\infty$, the infinite size system.
Whereas finite temperature calculations give results periodic in imaginary time,
which have to be extrapolated, 
the new approach provides the $\beta=\infty$ ($T=0$) result directly.
With increased number of iterations, the infinite system results become available 
at larger distances both in spatial and in imaginary time direction.
A fit to the exponential decay $G(\tau)\sim e^{-\tau\Delta}$ of our data
directly provides estimates for the gaps
$\Delta=0.5059(4)$ at $N=2$ and $\Delta=0.19(1)$ at $N=4$,
consistent with previous results\cite{ladderresults}.
We also show results for $L=20$ and $\beta=\infty$ 
to exemplify the effect of finite size systems.

%
\begin{figure}[htb]
\center{
 \epsfig{file=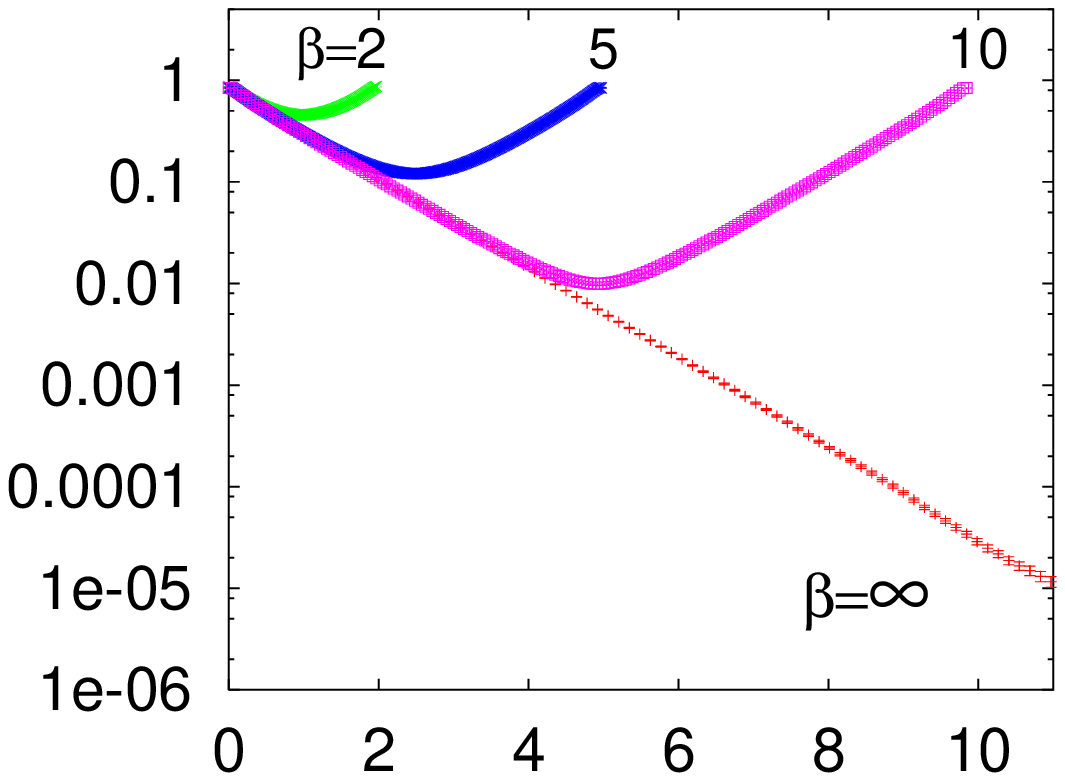,height=36mm}  
 \epsfig{file=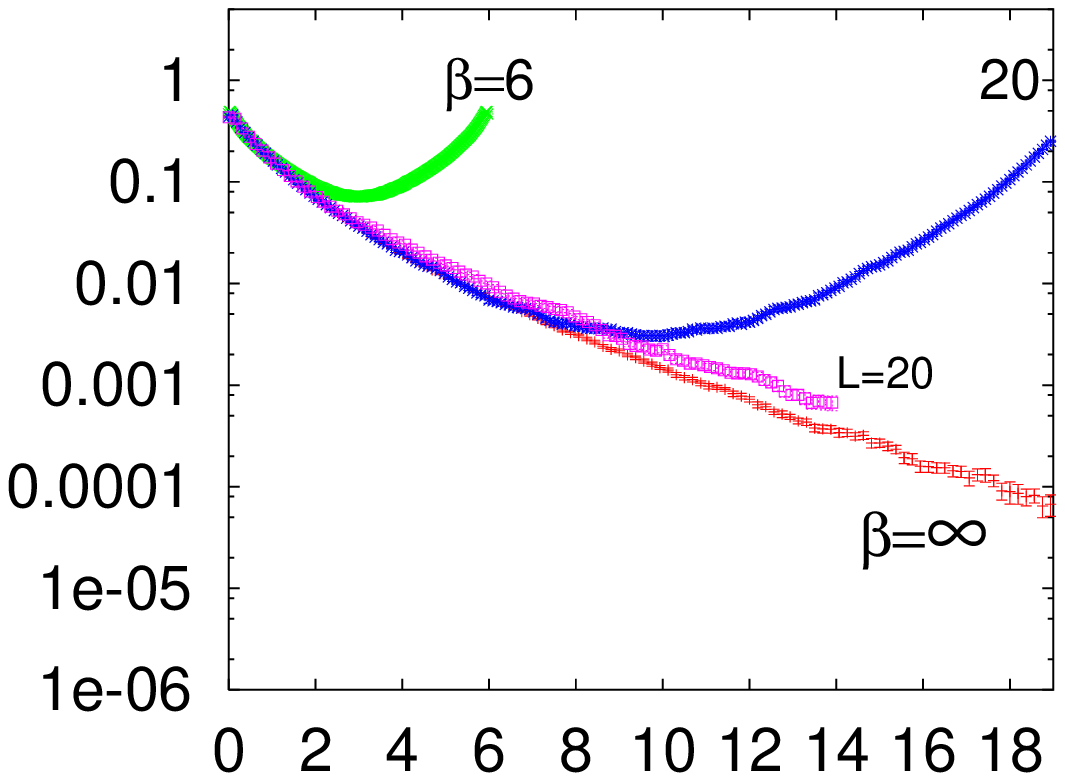,height=36mm}  
       }
     \caption{Greens functions 
              $\langle S(\vec{q},0) \, S(\vec{q},t) \rangle$
              at $\vec{q}=(\pi,\pi)$ for $L=\infty$, the infinite size system,
              with  $N=2$ (top) and $N=4$ (bottom).
              Results at $L=20$, $\beta=\infty$ have been added to exemplify finite size effects.
}\Label{temporal} 
\end{figure}
\begin{figure}[htb]
\center{
 \epsfig{file=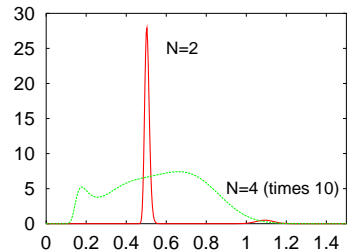,height=36mm}  
       }
     \caption{Spectrum $S(\vec{q},\omega)$ at $\vec{q}=(\pi,\pi)$ for $L=\infty$ and $\beta=\infty$
}\Label{spectrum} 
\end{figure}
Continuing the imaginary time greens function to real frequencies by the maximum entropy method
provides the spectra of \fig{spectrum}, in which the gaps, the single magnon peaks, 
and higher excitations for $N=4$ are clearly visible.
This appears to be the first time that the spectrum for $N=4$ has been calculated.


Let us compare our method to other approaches.
Exact diagonalization provides ground state results but is limited to small systems.
Projector methods are also limited in system size and rely critically on proper convergence
to the ground state.
DMRG can achieve $\beta=\infty$ (mostly without dynamical information)
for fairly small systems, or infinite size for large temperatures and finite $\Delta\tau$.
The most powerful method to extrapolate to infinite system size is 
the Finite Size Scaling method
of Kim \cite{Kim} and  Caracciolo et al.\ \cite{Caracciolo},
which allows extrapolation at correlation lengths far larger than the system size,
but requires knowledge about scaling and corrections to scaling of the model. 

In summary, we have introduced a new method,
as a small modification of existing cluster methods,
to simulate both classical and quantum systems at infinite system size
and/or at zero temperature, while obtaining all two-point functions
and derived quantities.
Larger distances (in space and/or imaginary time) are accessed for longer simulations.
The method in its present form
is applicable to all cases where existing single-cluster algorithms can be used,
as long as the unsubtracted two-point correlation function
decays sufficiently quickly,
including any $\beta<\beta_c$ in the unbroken phase.
Such points away from a transition are often the region of interest
when comparing simulation results to experimental measurements.





\begin{references}


\newcommand{\REF }[4]{     #1                   {#2}, #3 (#4)} 
\newcommand{\PRL }[3]{\REF{Phys. Rev.  Lett.\   }{#1}{#2}{#3}}
\newcommand{\PRB }[3]{\REF{Phys. Rev.\         B}{#1}{#2}{#3}}
\newcommand{\PRD }[3]{\REF{Phys. Rev.\         D}{#1}{#2}{#3}}
\newcommand{\PRE }[3]{\REF{Phys. Rev.\         E}{#1}{#2}{#3}}
\newcommand{\PLA }[3]{\REF{Phys. Lett.\        A}{#1}{#2}{#3}}
\newcommand{\NPB }[3]{\REF{Nucl. Phys.\        B}{#1}{#2}{#3}}
\newcommand{\ZPB }[3]{\REF{Z.    Phys.\        B}{#1}{#2}{#3}}
\newcommand{\CMP }[3]{\REF{Comm. Math. Phys.\   }{#1}{#2}{#3}}
\newcommand{\JPSJ}[3]{\REF{J. Phys. Soc. Jpn.\  }{#1}{#2}{#3}}
\newcommand{\JSP }[3]{\REF{J. Stat. Phys.\      }{#1}{#2}{#3}}
\newcommand{\etal}{et.\ al.}



\bibitem{SW} R.H. Swendsen and J.S. Wang, \PRL{58}{86}{1987}
             ;  for a review see e.g. 
                  A.D. Sokal, {\em Bosonic Algorithms}, 
                  in M. Creutz, ``Quantum Fields on the Computer'' (Advanced Series on Directions in High Energy Physics, Vol 11);
                  newer version available at http://www.dbwilson.com/exact/cargese.ps.gz.

\bibitem{FortuinKasteleyn} P.W. Kasteleyn and C.M. Fortuin, \JPSJ{26(Suppl.)}{11}{1969};
                           C.M. Fortuin and P.W. Kasteleyn, \REF{Physica}{57}{536}{1972}

\bibitem{EdwardsSokal} R.G. Edwards and A.D. Sokal, Phys. Rev. {\bf D38}, 2009 (1988).

\bibitem{Wolff}    U. Wolff, \PRL{62}{361}{1989}, \NPB{334}{581}{1990}

\bibitem{Caselle} M. Caselle and  F. Gliozzi, \REF{J.Phys}{A33}{2333}{2000};
                  M. Caselle, F. Gliozzi, and S. Necco, \REF{J.Phys}{A34}{351}{2001}

\bibitem{LoopAlg} H.G. Evertz, G. Lana, and M. Marcu, \PRL{70}{875}{1993},
                  for a review see H.G. Evertz, cond-mat/9707221 (2nd ed. Jan.\ 2000)

\bibitem{ladders} See e.g. E. Dagotto and M. Rice, \REF{Science}{271}{G18}{1996}

\bibitem{ladderresults} 
                       B. Frischmuth, B. Ammon, and M. Troyer, \PRB{54}{R3714}{1996};
                       S.R. White, R.M. Noack, and D.J. Scalapino, \PRL{73}{886}{1994}


\bibitem{conttime} B.B. Beard and U.J. Wiese, \PRL{77}{5130}{1996}


\bibitem{Kim} J.K. Kim, \PRL{70}{1735}{1993} ,\PRD{50}{4663}{1994}

\bibitem{Caracciolo} S. Caracciolo et al., \PRL{74}{2969}{1994} , \PRL{75}{1891}{1995}


\end{references}
\end{document}